\documentclass[onecolumn, numberedappendix,trackchanges]{aastex6}
\usepackage{graphicx}
\usepackage{gensymb}
\usepackage{epsfig}
\usepackage{natbib}
\usepackage{amssymb}
\usepackage{amsbsy}
\usepackage{natbib}
\usepackage{url}
\usepackage[mathcal]{euscript}
\bibpunct{(}{)}{,}{a}{}{,}

\newcommand{\mum}{$\mu$m}
\newcommand{\eg}{{\it e.g.}}
\newcommand{\etal}{et~al.}

\newcommand{\pas}{\ensuremath{{}^{\prime\prime}\mskip-8mu.\,}} 

\begin{document}
 
\def\simlt{\vcenter{\hbox{$<$}\offinterlineskip\hbox{$\sim$}}}
\def\simgt{\vcenter{\hbox{$>$}\offinterlineskip\hbox{$\sim$}}}
\def\etal{et al.\ }
\def\kms{km s$^{-1}$}

\title{Spitzer Space Telescope Mid-IR Light Curves of Neptune} 
\author{John Stauffer\altaffilmark{1}, 
Mark S. Marley\altaffilmark{2},
John E. Gizis\altaffilmark{3},
Luisa Rebull\altaffilmark{4,1}, 
Sean J. Carey\altaffilmark{1},
Jessica Krick\altaffilmark{1},
James G. Ingalls\altaffilmark{1},
Patrick Lowrance\altaffilmark{1},
William Glaccum\altaffilmark{1},
J. Davy Kirkpatrick\altaffilmark{5},
Amy A. Simon\altaffilmark{6},
Michael H. Wong\altaffilmark{7} 
}
\altaffiltext{1}{Spitzer Science Center (SSC), California Institute of
Technology, Pasadena, CA 91125, USA}
\altaffiltext{2}{NASA Ames Research Center, Space Sciences and
Astrobiology Division, MS245-3, Moffett Field, CA 94035 USA}
\altaffiltext{3}{Department of Physics and Astronomy, University of 
Delaware, Newark, DE  19716, USA}
\altaffiltext{4}{Infrared Science Archive (IRSA), 1200 E. California Blvd, 
MS 314-6, California Institute of Technology, Pasadena, CA 91125 USA}
\altaffiltext{5}{Infrared Processing and Analysis Center, MS 100-22, 
California Institute of Technology, Pasadena, CA 91125 USA}
\altaffiltext{6}{NASA Goddard Space Flight Center, Solar System
Exploration Division (690.0), 8800 Greenbelt Road, Greenbelt, MD 20771 USA}
\altaffiltext{7}{University of California, Department of Astronomy, Berkeley CA 94720-3411, USA}


\begin{abstract}

We have used the {\it Spitzer Space Telescope} in February 2016 to
obtain high cadence, high signal-to-noise, 17-hour duration light
curves of Neptune at 3.6 and 4.5 $\mu$m.  The light curve duration was
chosen to correspond to the rotation period of Neptune.  Both light
curves are slowly varying with time, with full amplitudes of 1.1 mag
at 3.6 $\mu$m and 0.6 mag at 4.5 $\mu$m.   We have also extracted
sparsely sampled 18-hour light curves of Neptune at W1 (3.4 $\mu$m)
and W2 (4.6 $\mu$m) from the {\it WISE/NEOWISE} archive at six epochs
in 2010-2015.  These light curves all show similar shapes and
amplitudes compared to the {\it Spitzer} light curves but with
considerable variation from epoch to epoch.   These amplitudes are
much larger than those observed with Kepler/K2 in the visible
(amplitude $\sim$0.02 mag) or at 845 nm with the {\it Hubble Space
Telescope} in 2015 and at 763 nm in 2016 (amplitude $\sim$ 0.2 mag).  
We interpret the {\it Spitzer} and {\it WISE} light curves as arising
entirely from reflected solar photons, from higher levels in Neptune's
atmosphere than for K2.  Methane gas is the dominant opacity source in
Neptune's atmosphere, and methane absorption bands are present in the
{\it HST} 763, and 845 nm, {\it WISE} W1, and {\it Spitzer} 3.6 $\mu$m
filters.

\end{abstract}


\section{Introduction}

Cloud formation in cool atmospheres is a key characteristic of
planetary, exoplanetary, and brown dwarf atmospheres
\citep{2013cctp.book..367M}.  We know from the study of isolated,
self-luminous, field brown dwarfs that these clouds will typically not
be completely uniform, but instead can have both vertical and
horizontal structure, and vary in time. Synoptic monitoring of objects
with patchy clouds is a powerful tool to study such objects because
each observation samples different regions of the surface as the
target rotates.  Spectroscopic monitoring of brown dwarfs with the
{\em Hubble Space Telescope} (HST) has detected wavelength dependent
phase shifts in light curve features \citep{buenzli2012, apai2013},
best interpreted as due to molecules condensing at different pressure
levels within the atmosphere and  cloud structures in the vertical as
well as horizontal \citep{buenzli2012, yang2016}, including in some
cases the presence of a low pressure haze layer \citep{yang2015}.  The
chemistry of the clouds varies with the temperature of the brown dwarf
photosphere, from the warmest condensates (e.g., various oxides of
magnesium and calcium) at $T_{\rm eff}$ $\sim$ 2500K to water clouds
for the coolest brown dwarfs \citep{faherty2014}.  Using the {\it
Spitzer Space Telescope} \citep{werner2004} to obtain mid-infrared
time series photometry, \citet{metchev2015} found that discrete cloud
spots are present in most L and T-type brown dwarfs, typically
producing variability in the range 0.2\% to 5\%. Even higher amplitude
variability is a characteristic of the L/T transition
\citep{radigan2014} where cloud properties change dramatically in a
narrow temperature range. Cloud lifetimes apparently vary from hours
\citep{gillon2013, metchev2015} to years in some cases
\citep{gizis2015}. \citet{cushing2016} found that a $\sim400$K Y dwarf
was variable at the 3.5\% level, with the clouds changing in the
months between observing epochs.  

Extending this work to planetary-mass objects orbiting brighter
primaries is challenging, but cloud variability in 2M1207b has been
detected \citep{zhou2016} and the HR 8799 planets are being monitored
\citep{apai2016}. Extensive studies of this nature likely must await
the James Webb Space Telescope or other future technologies.  
However, extending this work to the gas giants of the Solar System is
possible now.  And, while the surfaces of brown dwarfs and exoplanets
cannot be spatially resolved, imaging of the gas giants in the Solar
System is possible and can provide ground truth for inferences drawn
from synoptic monitoring.  Therefore, synoptic monitoring observations
of Solar System planets as ``point sources" provide a useful
comparison to these brown dwarf and exoplanet results, as well as
offering new insights into weather on the planets themselves. Although
there is an exceptional long-term optical $b$ and $y$ band photometric
sequence for Neptune \citep{lockwood1991} exhibiting variations in
reflected light of a few percent over time, until fairly recently
there has been little in the way of intensive, short term photometric
monitoring of Solar System giant planets reported in the literature.
Most recently \citet{srom2001} monitored Neptune over about one
rotation period in 1996 from HST and measured variability as large as
22\% at wavelengths with strong methane absorption. They attributed
the variability to patchy, high altitude clouds. Much longer term
cloud morphology changes for Neptune--over years--can be found in
\citet{karkoschka2011} and \citet{irwin2016}, who report that
near-infrared high resolution imaging detects significant cloud
evolution between 2009 and 2013. 

The repurposed Kepler/K2 Mission \citep{howell2014} notably altered
the status quo by providing the opportunity to obtain a continuous,
months-long lightcurve of Neptune in a single broad optical band (0.43
$\mu$m to 0.90$\mu$m). Neptune was observed by K2 in late 2014 at a
one-minute cadence for 49 days (Rowe \etal\ 2016, in prep). This K2
broadband optical light curve of Neptune is sensitive to light
reflected from the planet and not thermal emission from the
atmosphere. Thus the measured time-varying lightcurve is particularly
sensitive to the scattering properties and altitudes of clouds within
the atmosphere and the relative altitude of the clouds compared to
other sources of atmospheric opacity, primarily methane absorption. 

\citet{simon2016} analyzed the cloud contributions to the complex
Neptune K2 light curve. The light curve has an amplitude of $\sim 2\%$
and is dominated by discrete cloud features which are more (or less)
reflective than the global cloud deck. Because of Neptune's
latitude-dependent high winds, periods from 15 to 19 hours are
detected instead of a single rotation period. The most significant
periodic ($P=16.8$ hrs) signal is attributed to a single, long-lived
stable feature at a latitude of 45{\degree} S. This bright cloud was
confirmed by contemporaneous near-infrared Keck Adaptive Optics
imaging and later HST optical imaging. Contributions from smaller,
rapidly evolving discrete clouds add irregularities to the dominant
signal.

At Jupiter, the transition from reflected sunlight dominating the
disk-averaged emission to thermal emission takes place around $4\,\rm
\mu m$. The planet's famed five-micron hot spots, for example, are
regions of thermal emission from the deep atmosphere pouring out
through holes in the water and ammonia cloud decks in a region of low
gas opacity. Large horizontal inhomogeneity at $5\,\rm \mu m$
corresponds to variability in the tropospheric weather layer. At the
much colder Uranus and Neptune, the transition from scattered to
emitted light takes place at longer wavelengths, at about 6 to $7\,\rm
\mu m$, and the thermal emission originates from the higher-altitude,
radiatively controlled stratosphere. Examples of ice giant variability
studies at thermal wavelengths include the work of \citet{hammel2007}
who detected a bright polar region at 7.7 and 11.7 $\mu$m in Neptune,
and showed that these stratospheric ethane and methane emission
features are uncorrelated with the deeper cloud features observed in
near-infrared imaging. Based on independent, mid-IR ground-based
imaging, \citet{orton2007} detected a distinct rotating hot spot in
ethane and methane emission; \citet{orton2012} showed that this
feature has been absent in some years and may be a wave feature
excited by dynamics lower in the atmosphere.

In this paper, we provide the first well-sampled mid-infrared (mid-IR)
light curves for Neptune based on thirty-five hours of continuous
monitoring in early 2016 with  {\it Spitzer}. We also provide six
sparsely sampled mid-IR light curves of Neptune obtained in 2010
through 2015 with the Wide-field Infrared Survey Explorer \citep[{\it
WISE};][] {wright2010} satellite.  While these light curves are also
sensitive to reflected light, they probe different depths into
Neptune's cloudy atmosphere than those obtained previously with K2 and
HST. Our data thereby add new constraints on the composition and
structure or Neptune's cloud deck.  Section 2 describes the various
sources of data we use; \S3 and \S4 describe in detail the analysis
steps used to construct the {\it Spitzer} and {\it WISE} light
curves.  Section 5 compares the infrared light curves to those
obtained at shorter wavelengths and the initial inferences we draw
from these comparisons.

\section{Observational Data }

\subsection{{\it Spitzer} Data}
Neptune was observed between UT 2016 February 21-23 in both of the 3.6
$\mu$m (IRAC-1) and 4.5 $\mu$m (IRAC-2) channels of the Infrared Array
Camera \citep[IRAC;][]{fazio2004} on {\it Spitzer}.  The measurements
were part of Director's Discretionary Time Program 12125 (PI:
Stauffer).  

The time period was chosen expressly because Neptune as seen from {\it
Spitzer} was near the stationary point in its orbit, thus minimizing
the planet's motion during the many hour observation and thereby
minimizing any possible time-varying contamination from background
field stars that would pass through the Neptune or sky aperture for
the photometry.

The Astronomical Observation Requests (AORs) were made in IRAC's
staring mode, where for each channel, the spacecraft is maneuvered so
that the target is placed on the well-calibrated peak-up pixel and
back-to-back frames taken for the total time of the AOR with no
dithering.  For each channel, the total duration of the AOR was set to
cover a complete rotation of Neptune, or about 17.2 hours.   In
channel 1 (3.6 $\mu$m), frames with times of 100 seconds were used
(corresponding to 96.8 second exposure times), resulting in 622
images; in channel 2 (4.5$ \mu$m), a frametime of 30 seconds was used
(corresponding to 26.8 sec exposure times), resulting in 2018 images.
The image files were dark-subtracted, linearized, flat-fielded, and
calibrated using the S19.2 version of the IRAC pipeline. We had
requested that the channel 2 observations be made immediately
following the channel 1 observations, but a time -critical exoplanet
transit observation was inserted between the two Neptune AORs,
resulting in the channel 2 light curve beginning about 2.3 days after
the start of the channel 1 observation. 

Flux densities were measured with aperture photometry on the {\it
Spitzer} Basic Calibrated Data (BCD) images.  After first determining
the center of Neptune via a ``center of light'' technique
\citep{ingalls2016}, the flux was integrated over a 5 pixel radius
aperture using the IDL Astronomy User's Library routine {\tt
aper.pro}\footnote{http://idlastro.gsfc.nasa.gov/ftp/pro/idlphot/aper.pro}.
The per-pixel average background  was derived for each image using an
annulus centered on Neptune with inner and outer radii of (16,18)
pixels.  This per-pixel bias was scaled to the aperture area and
subtracted from the integrated flux.  Normally, a tighter annulus is
used to derive the background for point source photometry but avoiding
Triton and a background star required a minimum radius of 16 pixels. 
We corrected the fluxes for losses due to a finite aperture size by
multiplying by the aperture corrections:  1.049 at 3.6 $\mu$m and
1.050 at 4.5 $\mu$m.  These corrections were derived for point source
observations with post-cryogenic IRAC for an aperture of 5 pixels
radius and a background annulus of (10,20) pixels.  Simulations with
the IRAC Point Response Function
(PRF)\footnote{http://irsa.ipac.caltech.edu/data/SPITZER/docs/irac/calibrationfiles/psfprf/}
show that using a nonstandard (16,18) background annulus changes the
flux by only 0.2\% compared to the (10,20) annulus. However, these
aperture corrections were not optimized for non-point objects
(Neptune's angular diameter from Earth ranges from 2\pas 17 to 2\pas
37, whereas the IRAC point spread function (PSF) at 3.6 and 4.5 $\mu$m
is about 1\pas 8 FWHM).  We expect therefore some unaccounted-for
aperture losses which do not affect relative flux estimates. 
Additional losses due to the under-sampled PSF and a variable spatial
response of each pixel (the so-called ``pixel-phase effect''), {\it
can} affect relative flux variations when coupled with normal
spacecraft pointing fluctuations.  However, the magnitude of this
effect is only a few percent \citep[about one tick mark in Figure
\ref{fig:SpitzerLC}; see][]{2012SPIE.8442E..1YI}  for point sources
and less for partially resolved sources like Neptune, and so we did
not attempt to correct for it here.

We converted aperture fluxes to magnitudes using the in-band flux
densities of Vega:  278\,Jy (3.6 $\mu$m) and 180\,Jy (4.5 $\mu$m).  
The fluxes reported for both IRAC and WISE are not color corrected for
the difference between the spectral energy distribution of Neptune and
the fiducial spectral shape $F_{\nu} \propto \nu^{-1}$ used in the
calibration of IRAC \citep{reach2005}, or the $F_{\nu} \propto
\nu^{-2}$ adopted for the calibration of WISE \citep{wright2010}.  Not
attempting a color correction should not affect the measured color
variation unless the underlying spectral energy distribution is also
varying significantly.  For a detailed color comparison between the
IRAC and WISE data, a SED must be assumed and color corrections
applied to account for the different bandpass profiles and the
different calibration assumptions.

\subsection{{\it WISE} Data}

{\it WISE} was launched on 2009 December 14 to survey the sky in four
broad wavelength bands referred to as W1 (3.4 $\mu$m), W2 (4.6
$\mu$m), W3 (12 $\mu$m), and W4 (22 $\mu$m). After an in-orbit
checkout, {\it WISE} began surveying the sky on 2010 January 14 in a
Sun-synchronous polar orbit around the Earth, allowing the entire sky
to be mapped in just six months. Along the ecliptic, each location is
visited a minimum of eight times during each sky pass, and each of
those visits has an exposure time of 7.7s at W1 and W2 and 8.8s at W3
and W4. Those data are generally confined to a $\sim$24-hour period
since the observatory sweeps through roughly one degree in ecliptic
longitude per day. Exceptions to this are when the scan track is
temporarily toggled to either larger or smaller longitudes in order to
avoid the boresight's passing within five degrees of the Moon. In such
cases, observations of a particular sky location may be interrupted by
a couple of days while the Moon passes, in which case the coverage at
that epoch may be spread over a few days surrounding the gap. 

{\it WISE} completed its first full pass of the sky on 2010 July 17.
On 2010 August 05, during the second pass, the outer, secondary tank
was depleted of its cryogen, rendering the W4 band unusable. This was
followed by depletion of cryogen from the inner, primary tank on 2010
September 30, which rendered the W3 band unusable. The two
short-wavelength bands were largely unaffected, so WISE continued to
collect data in W1 and W2 until the spacecraft was placed into
hibernation on 2011 January 31, having then completed two full sky
passes and 20\% of a third. {\it WISE} observations resumed in 2013
December as the re-branded {\it NEOWISE} mission, which is scheduled
to map the sky six additional times. {\it NEOWISE} observations will
continue through the end of 2016.

The W1 band has a central wavelength designed to fall in the middle of
the strong 3.3 $\mu$m fundamental methane absorption band seen in
colder brown dwarfs and giant planets. The W2 band has a central
wavelength of 4.6 $\mu$m, where the atmospheres of these objects are
relatively transparent to radiation, assuming a cloud-free atmosphere.
An example is the 5-$\mu$m region in the atmosphere of Jupiter, where
thermal emission can be seen through holes in the thick cloud layer
(\citealt{westphal1969}). These two bandpasses therefore can sample
two very different physical regimes in the atmospheres of these
objects and can be used to study their chemistry and physics (see
Figure 2 of \citealt{mainzer2011}). 

As can be gleaned from the {\it WISE} bandpass coverages shown in
Figure 7 of \cite{wright2010} and the {\it Spitzer}/Infrared
Spectrograph (IRS) spectra of known brown dwarfs in Figure 5 of
\cite{cushing2006}, the W3 band samples a wavelength regime including
CH$_4$ and NH$_3$ absorption troughs. The W4 band, however, samples a
region of the Rayleigh-Jeans tail with no sharp, deep molecular
absorption bands, at least as predicted by brown dwarf atmospheric
models (\citealt{burrows2003}).

Neptune has been observed at six different epochs in the currently
available {\it WISE} and {\it NEOWISE} data releases. Its magnitudes
range from W1$\approx$10.1-11.5 mag, W2$\approx$9.5-11.1 mag,
W3$\approx$2.8-3.1 mag, and W4$\approx$0.0-0.3 mag. As stated in the
{\it WISE} All-Sky Release Explanatory Supplement
(\citealt{cutri2012}), the detectors saturate at W1, W2, W3, and W4
magnitudes of 8.1, 6.7, 3.8, and $-$0.4 mag, respectively. The
profile-fit photometric measurements can extend the usable range to
brighter magnitudes by using the non-saturated wings of the stellar
profile, and this extends {\it WISE} photometric measurements to W1,
W2, W3, and W4 levels of 2.0, 1.5, $-$3.0, and $-$4.0 mag,
respectively. Because the measured magnitudes for Neptune are fainter
than these limits -- and because the apparent size of Neptune
($\sim$2{\farcs}2) is far less than the {\it WISE} resolution
(6{\farcs}1, 6{\farcs}4, 6{\farcs}5, and 12{\farcs}0 in W1 through W4,
respectively) -- the profile-fit photometry listed for Neptune in the
{\it WISE} Single Exposure Source Data Bases should be of good
quality.

\subsection{HST Imaging}

HST is capable of providing images of with spatial resolution of
Neptune's atmosphere $\sim$1000-2000 km, thus providing dozens of
resolution elements over Neptune's 50,000 km diameter. Although HST
imaging was not obtained at the same time as the {\it Spitzer} data,
global Neptune maps were obtained in 2015 September as part of the
Hubble Outer Planet Atmospheres Legacy (OPAL) program \citep[][Figure
5]{simon2016}.  These maps were obtained in multiple filters, allowing
direct viewing of the current cloud features, but with coarse temporal
spacing.  All filters redward of 600 nm show similar cloud features,
with good contrast against the background atmosphere.  The filter used
most often, F845M, provides the best light curve for comparison with
other data sets.  A discrete storm system was visible throughout 2015
near $45^\circ$ S and dominated the Hubble (16$\%$ peak-to-trough
variation at 845 nm) and 49-day Kepler lightcurves (2$\%$
peak-to-trough variation) \citep{simon2016}.  

This bright cloud feature at 45{\degree} S was suspected to be a
bright companion cloud to an unseen dark vortex, similar to the Great
Dark Spot seen by Voyager 2 \citep{smith1989}. The 2015 HST
observations indeed detected a dark feature at blue wavelengths, which
was confirmed by more recent Hubble imaging acquired on 2016 May 15
and 16 \citep{wong2016}. The bright companion clouds were again
present in the 2016 observations, strongly suggesting that they were
still present during the {\it Spitzer} observations, though at that
time, Neptune was too close to the Sun for ground-based verification.
Another large bright storm system was also present in the 2016 data
between 15 and $45^\circ$ S and was the brightest feature on the
planet. This data set was not optimized for temporal sampling, with
only 4 views over the course of 27.5 hours.  At 763 nm, Neptune again
showed variations of about 0.2 mag due to the larger cloud system.

\section{{\it Spitzer} Light Curves}

\begin{figure}[ht]
\centering
\includegraphics[width=9.5cm]{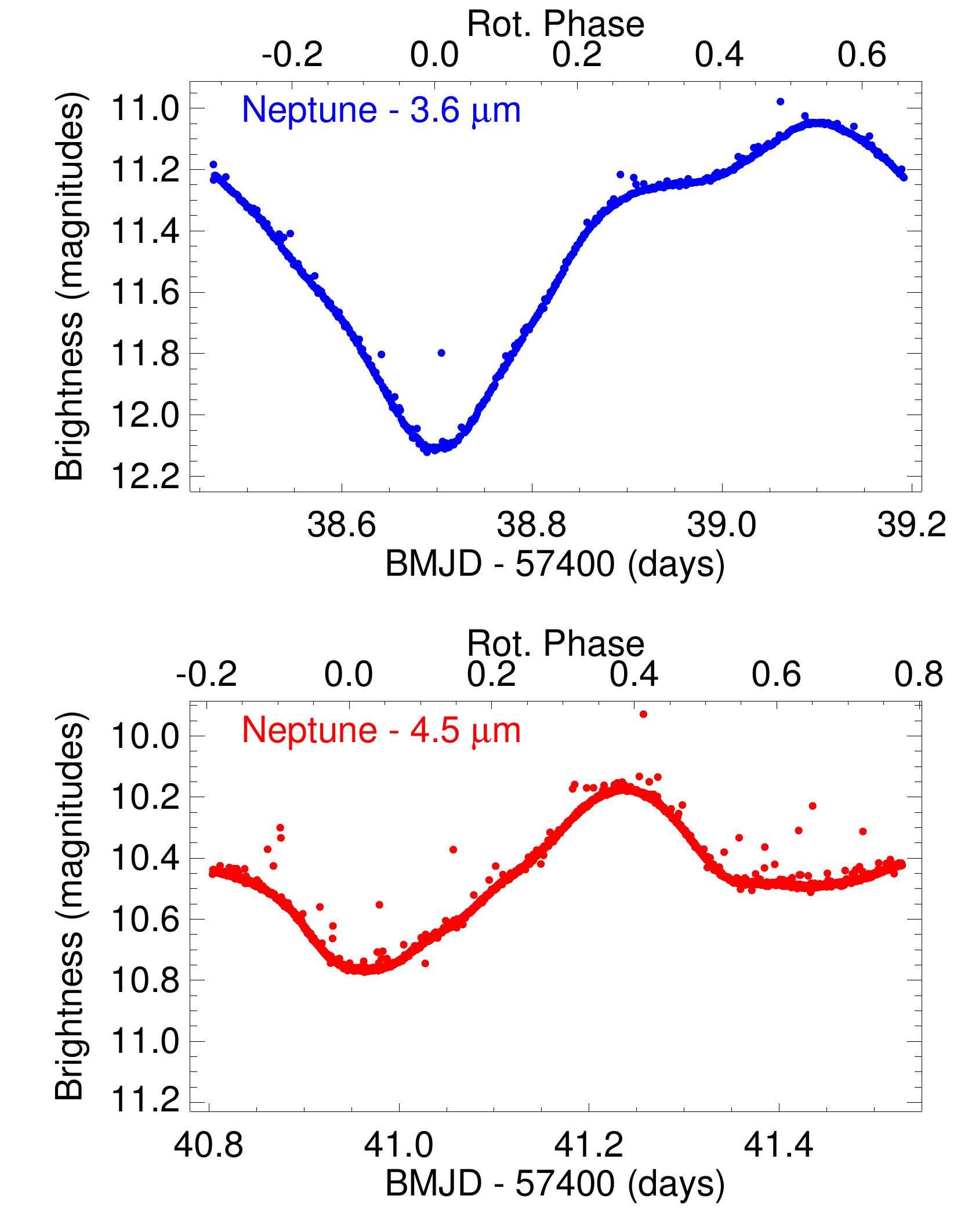}
\caption{{\it Spitzer} light curves of Neptune from 2016 February, in
the 3.6 $\mu$m ({\it top}) and 4.5 $\mu$m ({\it bottom}) bands of
IRAC.  Vertical scales of both plots cover a range of 1.4 mag.  Upper
axes indicate rotational phase, derived using a period of 18\,hr and
defining zero phase at  the time of the 3.6\,\micron\ minimum.
\label{fig:SpitzerLC}}
\end{figure}

We display light curves of Neptune measured with {\sl Spitzer}/IRAC in
Figure \ref{fig:SpitzerLC}.   The light curve data are provided in
tabular form in Table~\ref{tab:Spitzer_3.6} and 
Table~\ref{tab:Spitzer_4.5}. These are the first continuous Neptune
light curves  covering a full rotation at mid-IR wavelengths.  The
amplitude at 3.6 $\mu$m is  1.06\,mag, corresponding to a
peak-to-trough flux variation of more than a factor of 2.5; at  4.5
$\mu$m the amplitude is 0.59\,mag, corresponding to a peak-to-trough
flux variation of about 1.7.   These are consistent with the
amplitudes measured in the similar passbands {\it WISE} W1 and W2, 
respectively (see below), but are much larger than the amplitudes of
the K2 0.65\,\micron\ 
lightcurve of 2015 January \citep[$\sim 2\%$;][]{simon2016}, as well as the HST 845\,nm measurements (16\%). 

We label the upper axes of Figure~\ref{fig:SpitzerLC} in terms of
rotational ``phase,'' derived using a rotation rate of 18.0\,hr,
corresponding to zonal winds at $\sim 25\arcdeg$ latitude
\citep{smith1989, hammel1997}.  Assuming these observations bear the
imprint of the brightest feature seen in the 2016 May HST images, an
18\,hr period is the approximate mean of the two largest power
spectral density (PSD) peaks found in K2 data corresponding to zonal
winds between $20\arcdeg$ and $30\arcdeg$ latitude \citep[][Fig.
2]{simon2016}.  We define zero phase to occur at the approximate time
of the 3.6 $\mu$m minimum, defined as the barycentric modified Julian
date (BMJD) at the start of the 100\,s 3.6\mum\ integration with the
lowest flux, BMJD = 57438.697536 days.  As seen in Fig.
\ref{fig:SpitzerLC}, the 4.5\,$\mu$m light curve obtains its minimum
about 54 hours after the 3.6\,$\mu$m curve, which is near phase zero
under this definition.  In other words, the 3.6 $\mu$m and 4.5 $\mu$m
light curves exhibit minimum reflectance with a timing that is
consistent with the rotational period of zonal winds at $\sim
25\arcdeg$ latitude.   

\clearpage

\begin{deluxetable}{lcc}
\tabletypesize{\scriptsize}
\tablecolumns{3}
\tablewidth{0pt}
\tablecaption{{\it Spitzer} 3.6 $\mu$m Light Curve for Neptune\tablenotemark{a}\label{tab:Spitzer_3.6}}
\tablehead{
\colhead{BMJD\tablenotemark{b}}  & \colhead{[3.6]} & \colhead{Uncertainty} \\
\colhead{days} & \colhead{mag} & \colhead{mag} 
 }
\startdata
57438.46455 &11.1835 & 0.0063 \\
57438.46470 & 11.2340 & 0.0018 \\
57438.46587 &11.2193 & 0.0018 \\
57438.46704 & 11.2196 & 0.0020 \\
57438.46821 & 11.2234 & 0.0020 \\
57438.46938 & 11.2256 & 0.0020 \\
57438.47055 & 11.2304 & 0.0020 \\
57438.47172 & 11.2355 & 0.0019 \\
57438.47288 & 11.2376 & 0.0019 \\
57438.47406 & 11.2398 & 0.0018 \\
\enddata
\tablenotetext{a}{This table is available in its entirety in the online
version. A portion is shown here to demonstrate its form and content.}
\tablenotetext{b}{Barycentric Modified Julian Date at the start of the 100\,s integration.}
\end{deluxetable}

\begin{deluxetable}{lcc}
\tabletypesize{\scriptsize}
\tablecolumns{3}
\tablewidth{0pt}
\tablecaption{{\it Spitzer} 4.5 $\mu$m Light Curve for Neptune\tablenotemark{a}\label{tab:Neptune4.5}\label{tab:Spitzer_4.5}}
\tablehead{
\colhead{BMJD}  & \colhead{[4.5]} & \colhead{Uncertainty} \\
\colhead{days} & \colhead{mag} & \colhead{mag} 
 }
\startdata
57440.80363 & 10.4527 & 0.0026 \\
57440.80399 & 10.4387 & 0.0026 \\
57440.80435 & 10.4371 & 0.0026 \\
57440.80471 & 10.4451 & 0.0026 \\
57440.80507 & 10.4385 & 0.0026 \\
57440.80543 & 10.4420 & 0.0025 \\
57440.80579 & 10.4441 & 0.0025 \\
57440.80615 & 10.4416 & 0.0026 \\
57440.80651 & 10.4423 & 0.0026 \\
57440.80687 & 10.4431 & 0.0026 \\
\enddata
\tablenotetext{a}{This table is available in its entirety in the online
version. A portion is shown here to demonstrate its form and content.}
\tablenotetext{b}{Barycentric Modified Julian Date at the start of the 30\,s integration.}
\end{deluxetable}

\section{{\it WISE} Light Curves}

\begin{figure}
\centering
\includegraphics[scale=0.65,angle=90]{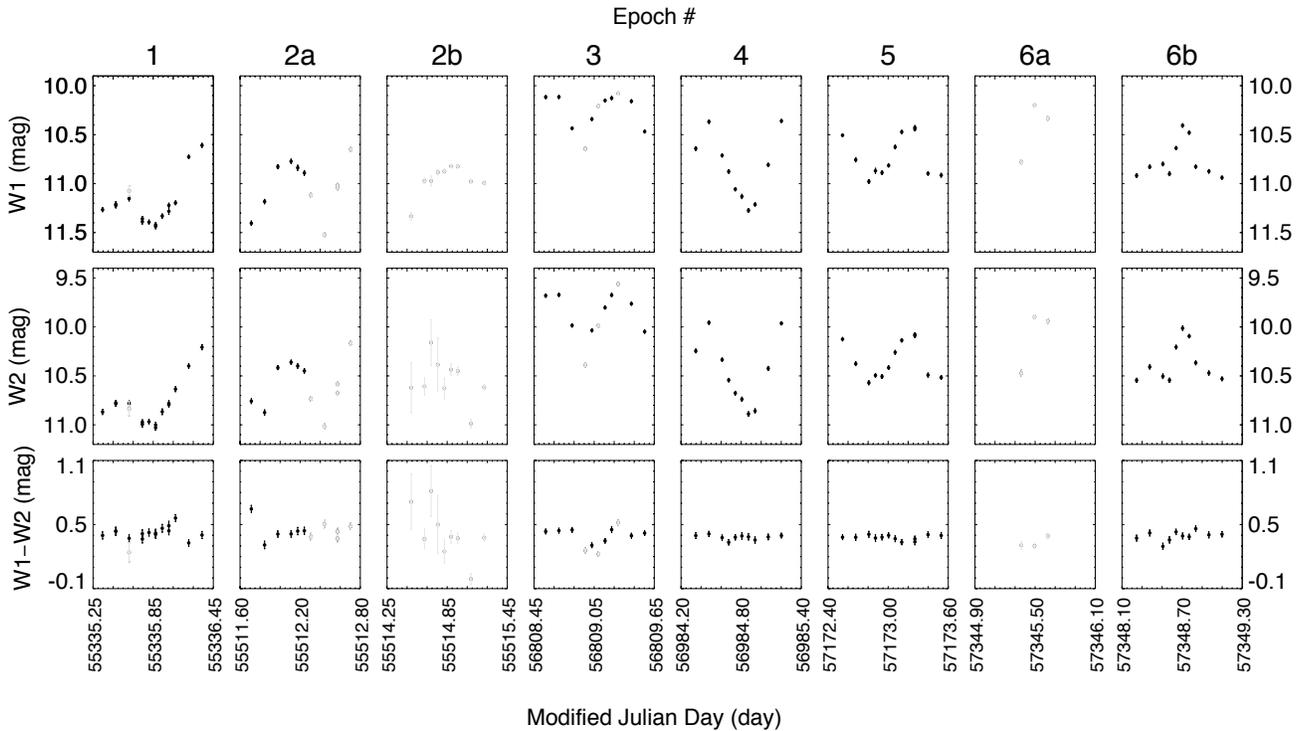}
\caption{Light curves of Neptune using {\it WISE} W1 and W2 data. The top row shows the variation in W1 mag, the second row shows W2 mag, and the bottom row shows the W1$-$W2 color. Each column represents a different epoch, as defined in Table~\ref{tab:WISE_epochs}. For many of the data points shown, the symbol size is larger than the photometric uncertainties, so error bars are not visible. Points in black are those used in the statistics computed in Table~\ref{tab:WISE_epochs}. Points in light gray are ones that failed to meet our criteria for high-quality photometry. See text for details.
\label{fig:WISE_W12_light_curves}}
\end{figure}

Light curves at the {\it WISE} W1 and W2 bands are shown in
Figure~\ref{fig:WISE_W12_light_curves}. The {\it WISE} data are
provided in tabular form in Table~\ref{tab:WISE}. This photometry can
be found in the Single Exposure Source Tables from the All-Sky,
Post-Cryo, and NEOWISE-R periods available through the NASA/IPAC
Infrared Science Archive\footnote{\url{http://irsa.ipac.caltech.edu}}
(IRSA). As shown in the figure, there are six epochs of data. For the
second and sixth epochs, the usual $\sim$24-hr observing window was
interrupted by a $\sim$2-day moon toggle, so data have been split into
``a'' and ``b'' subgroups. 

\begin{deluxetable}{lcccc}
\tabletypesize{\scriptsize}
\tablecolumns{5}
\tablewidth{0pt}
\tablecaption{{\it WISE} Light Curve Data for Neptune\tablenotemark{a}\label{tab:WISE}}
\tablehead{
\colhead{MJD}  & \colhead{W1} & \colhead{W1 Uncertainty} & \colhead{W2} & \colhead{W2 Uncertainty} \\
\colhead{days} & \colhead{mag} & \colhead{mag} & \colhead{mag} & \colhead{mag}
 }
\startdata
  55335.345131&  11.266&      0.020&  10.868&      0.026 \\
  55335.477308&  11.211&      0.027&  10.775&      0.024 \\
  55335.477435&  11.223&      0.025&  10.784&      0.026 \\
  55335.609612&  11.074&      0.052&  10.834&      0.072 \\
  55335.609739&  11.153&      0.023&  10.781&      0.025 \\
  55335.741916&  11.359&      0.024&  10.995&      0.027 \\
  55335.742043&  11.388&      0.024&  10.973&      0.025 \\
\enddata
\tablenotetext{a}{This table is available in its entirety in the online
version. A portion is shown here to demonstrate its form and content.}
\end{deluxetable}


We have performed quality analysis on the photometry.  Unless
otherwise stated, the data appear to be excellent.  Where there are
apparent issues, we discuss them in order of their epoch (where epoch
numbers appear in Figure~\ref{fig:WISE_W12_light_curves} and below in
Table~\ref{tab:WISE_epochs}):

\begin{itemize}

\item Epoch 1: The observation at MJD 55335.609 (specifically, the one
with scan/frame ID 04717b151) has Neptune very close to an array edge,
and the photometric uncertainties are correspondingly much higher than
normal. This point has been removed from further analysis. 

\item Epoch 2a: The last five observations at this epoch have moon
mask flags indicating that they fall within the zone where moonlight
can severely contaminate the W2 backgrounds; the last three
observations have this same moon mask flag set for W1. A check of the
images at these dates confirms high backgrounds in both bands. As a
result, we have removed all five of these data points from subsequent
analysis. 

\item Epoch 2b: All of the data at this epoch are likely contaminated
by moonlight in both W1 and W2, as indicated by the moon mask flag.
The high photometric uncertainties for these points as well as a
visual check of the images indicate that moonlight is indeed quite
severe for the majority of these observations. All data at this epoch
have been removed from subsequent analysis. 

\item Epoch 3: The data for MJD values of 56808.960 (scan/frame ID
49349b151), 56809.092 (49353b151), and 56809.289 (49360a126) have
image quality flags indicating streaked frames. These data points have
been removed from further analysis. 

\item Epoch 6a: All three data points have potential problems with
high backgrounds due to moonlight, as indicated by the moon mask flag.
The images confirm that the moonlight in these frames is severe. All
three of these points have been removed from subsequent analysis.

\end{itemize}

For each epoch, Table~\ref{tab:WISE_epochs} summarizes the mean
magnitude and the amplitude of the variability in each band and in the
[W1]$-$[W2] color. We note that the mean W1 and W2 magnitudes have
changed by $\sim$1.0 mag between epoch 1 and epoch 3. Variations this
large are sometimes also seen within an individual epoch, such as in
epochs 1 and 4, with variations changing gradually on timescales
similar to the {\it WISE} sampling of 90 minutes
(Figure~\ref{fig:WISE_W12_light_curves}). 

Despite the large excursions in W1 and W2 magnitude over the course of
a single day, the [W1]$-$[W2] color remains relatively constant at
$\sim$0.4 mag, although some excursions from this value, such as for
the first data point at epoch 2a, are seen. This same value of
[W1]$-$[W2] $\approx$ 0.4 mag also holds across the entire 5.5-yr data
set.

Simultaneous data at W3 and W4 also exist for epoch 1. At W3 band, the
core of Neptune's PSF is saturated, and the reduced $\chi^2$ values
from the PSF-fit photometry are generally well above 1.0, indicating
that the PSF fit was a poor match to the observed profile. Even though
the W4 magnitude is fainter than the saturation limit, the reduced
$\chi^2$ values from the PSF fit are also much greater than 1.0. As a
result, the quoted photometric errors in both bands are likely
underestimated given that the photometric measurements themselves are
likely biased. Since this makes interpretation of any real variability
difficult, we quote only the mean W3 and W4 magnitudes in
Table~\ref{tab:WISE_epochs}.

\begin{deluxetable*}{lccccccccccc}
\tabletypesize{\scriptsize}
\tablecolumns{12}
\tablewidth{0pt}
\tablecaption{{\it WISE} Neptune Observation Log\label{tab:WISE_epochs}}
\tablehead{
\colhead{Epoch} & 
\colhead{MJD} & 
\colhead{\# of} &
\colhead{\# of} &
\colhead{Mean} &
\colhead{Mean} & 
\colhead{Mean} & 
\colhead{Mean} & 
\colhead{W1} & 
\colhead{W2} & 
\colhead{Mean} &
\colhead{W1$-$W2} \\
\colhead{\#} & 
\colhead{Range} &
\colhead{Meas.} &
\colhead{Meas.} &
\colhead{W1} & 
\colhead{W2} & 
\colhead{W3} & 
\colhead{W4} & 
\colhead{Amp.} &
\colhead{Amp.} &
\colhead{W1$-$W2} &
\colhead{Amp.} \\
\colhead{} & 
\colhead{(day)} &
\colhead{Total} &
\colhead{Used} &
\colhead{(mag)} & 
\colhead{(mag)} & 
\colhead{(mag)} & 
\colhead{(mag)} & 
\colhead{(mag)} & 
\colhead{(mag)} & 
\colhead{(mag)} & 
\colhead{(mag)} 
}
\startdata
1 & 55335.3-55336.4& 16& 15&   11.21&   10.79&    3.02&    0.19&    0.83&    0.82&    0.42&    0.20\\
2a& 55511.7-55512.8& 11&  6&   10.99&   10.54& \nodata& \nodata&    0.63&    0.52&    0.44&    0.34\\
2b& 55514.4-55515.3&  9&  0& \nodata& \nodata& \nodata& \nodata& \nodata& \nodata& \nodata& \nodata\\
3 & 56808.5-56809.6& 11&  8&   10.24&    9.84& \nodata& \nodata&    0.35&    0.38&    0.41&    0.15\\
4 & 56984.3-56985.3& 10& 10&   10.84&   10.46& \nodata& \nodata&    0.91&    0.93&    0.38&    0.08\\
5 & 57172.5-57173.6& 12& 12&   10.72&   10.34& \nodata& \nodata&    0.55&    0.49&    0.38&    0.07\\
6a& 57345.3-57345.7&  3&  0& \nodata& \nodata& \nodata& \nodata& \nodata& \nodata& \nodata& \nodata\\
6b& 57348.2-57349.1& 10& 10&   10.76&   10.37& \nodata& \nodata&    0.53&    0.53&    0.39&    0.17\\
\enddata
\tablecomments{Epoch 1 is from the All-Sky phase of the mission, epoch 2 is from the Post-Cryo period, and epochs 3-6 are from the ongoing {\it NEOWISE} survey. Coverage of the field at epochs 2 and 6 was interrupted by a toggle of the spacecraft's scan path to avoid excessive moonlight.}
\end{deluxetable*}

\section{Physical Interpretation}

\begin{figure}[ht]
\centering
\includegraphics[scale=0.50,angle=0]{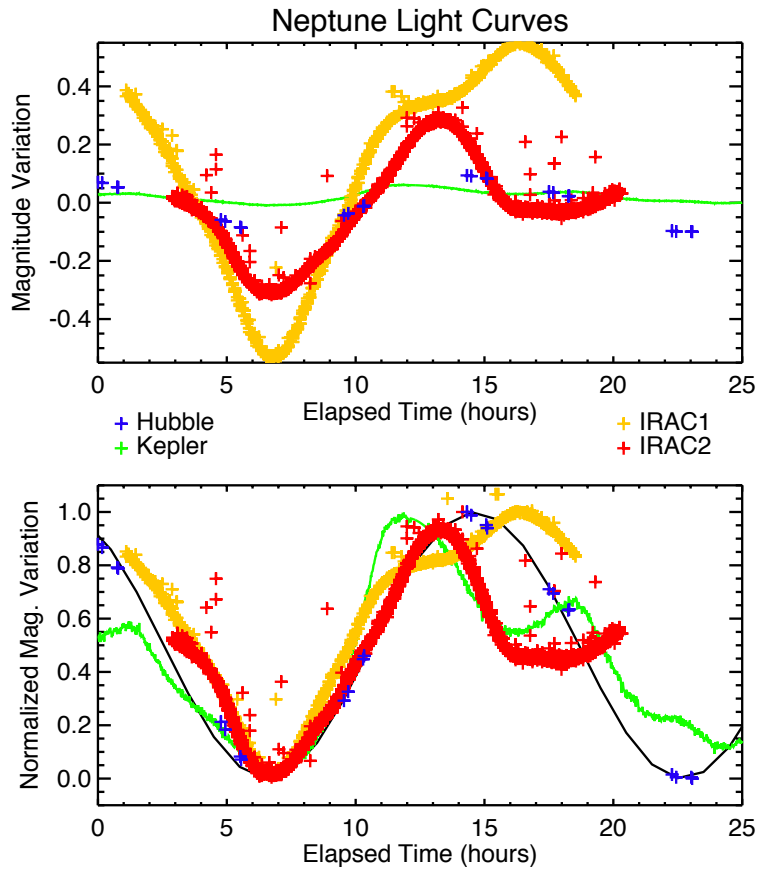}
\caption{Normalized magnitude light curves of Neptune using {\it
Hubble} 2015, {\it K2} and {\it Spitzer} data.  The top panel
shows the magnitude variations from each facility, time-shifted to
align their first minima and centered on their mean values, allowing
comparison of light curve amplitudes.  The bottom panel shows the
same, but with the amplitude of the variations normalized between 0
and 1 to allow comparison of light curve shapes. The solid black line
indicates Neptune's rotation period, with no adjustment for the
planet's inclination.}
\label{fig:All_light_curves}
\end{figure}

\begin{figure}[ht]
\centering
\includegraphics[scale=0.40,angle=0]{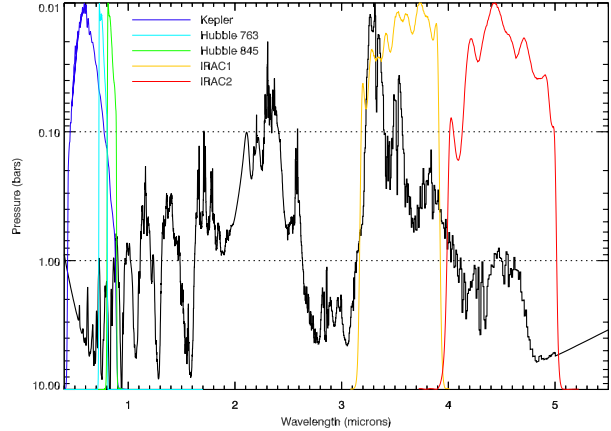}
\caption{Atmospheric transmission for Neptune or Uranus (P.\ Irwin,
personal communication).  The black line shows the approximate
transmission level for 50\% absorption in a cloud-free atmosphere,
including Rayleigh scattering, $\rm H_2-H_2$ and $\rm H_2-He$
collision-induced absorption, and methane gas absorption.  Filter
bandpasses are shown by the colored lines. In a cloud-free atmosphere,
each filter should be sensitive to different altitudes, with data near
$\sim$3 $\mu$m most sensitive to stratospheric levels. Cloud layers
overlying the indicated pressure level are visible in reflected light.
We see no evidence that the IR bandpasses sampled clouds at different
altitudes, which would manifest as different photometric amplitudes, 
a phase shift, or other significant morphological differences between
lightcurves.  }
\label{fig:transmission}
\end{figure}

The IRAC-1 (3.6 $\mu$m) and IRAC-2 (4.5 $\mu$m) light curves look very
similar to the light curves obtained by {\it K2} and {\it Hubble} in
2014/2015, despite the difference in the amplitude of the
variations, Figure~\ref{fig:All_light_curves}, top panel. 
Figure~\ref{fig:All_light_curves} bottom panel illustrates
this assertion, by over-plotting portions of the light curves from
each facility, arbitrarily phase shifting their $x$-axis
location so that the light-curve minima are aligned.  These
observations are nominally sensitive to a different altitude in the
atmosphere.  Figure~\ref{fig:transmission} shows that, in a cloud-free
atmosphere, each filter should be sensitive to different altitudes,
with the IRAC-1 band most sensitive to stratospheric levels, similar
to the {\it WISE} W1 (3.4 $\mu$m) filter.  

However, if clouds and hazes lie above the point in the atmosphere
where the column gas opacity becomes substantial (e.g., above the
black curve in Figure 4), they can dominate reflectance. This leads to
a flattening of the reflectance spectrum.  In spatially resolved
observations, limb and nadir observations can be combined to better
constrain cloud heights by leveraging the different atmospheric
columns \citep{simon1996}. Likewise, in the forward scattering realm
probed by exoplanet transit observations, the longer path lengths near
the limb allow even low-opacity hazes to strongly affect transmission
spectra \citep{fortney2005}, eliminating detectable gas absorption
bands in the atmospheres of, \eg, HD 189733b and GJ1214b
\citep{pont2008,berta2012,bean2011}. 

In our disk integrated mid-IR lightcurves, we see little difference
between the $\sim 3.6$ and $\sim 4.5\,\rm \mu m$ channels, despite the
large differences in methane opacity between the two bands. This
implies that any cloud opacity must lie sufficiently high in the
atmosphere that both bands see a comparable contribution. Judging from
Figure 4, this suggests that optically thick clouds, or at least the
cloudtops, must lie above about 0.30 bar. Further detailed modeling is
required to  refine this estimate.

Furthermore, we find little morphological differences between the two
lightcurves. Differences in cloud evolution or wind speed as a
function of altitude would manifest as different periods in the light
curves. A variation in the wind speed with altitude is expected based
on the 3D temperature field measured by Voyager
\citep{conrath1991,fletcher2014}. Recent Keck observations have
detected vertical windshear on Neptune by tracking resolved cloud
features at two different infrared wavelengths \citep{tollefson2016},
although the majority of discrete tracked features are too small and
short-lived to create periodic photometric signals in the short
duration lightcurve reported here. In general, the unresolved
observations are sensitive to very large cloud systems with
significant vertical range and uniform motion, producing similar
lightcurves at all wavelengths studied.

\begin{figure}[ht]
\centering
\includegraphics[scale=.4,angle=0]{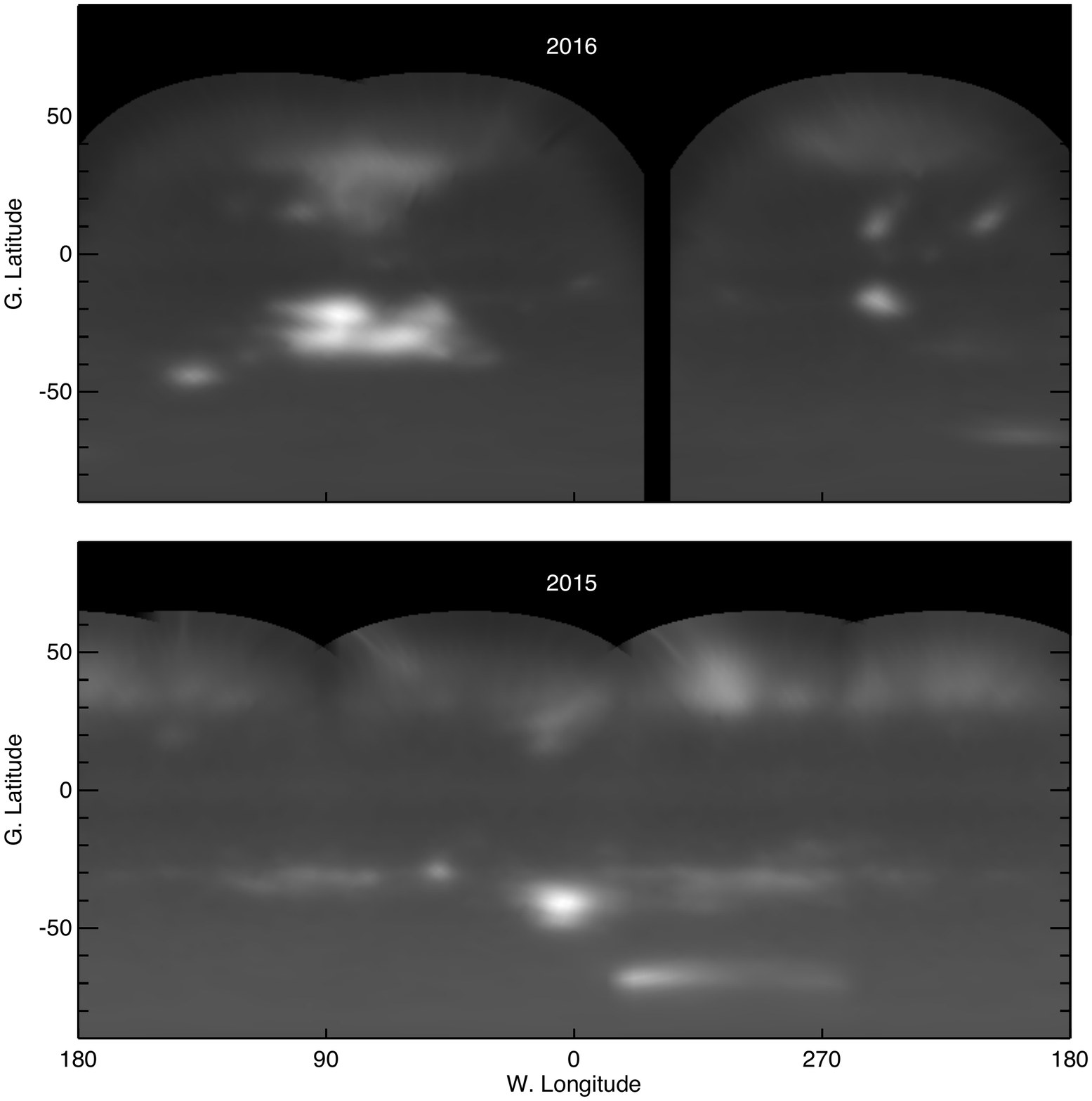}
\caption{Hubble maps of Neptune at 763 nm.  These data are rectilinear
mapped at 0.5 deg pix$^{-1}$ resolution and cover all latitudes and
longitudes.  Latitude coverage is incomplete due to Neptune's tilt. A
gap in the 2016 longitude coverage is produced because two opposite
hemispheres were imaged separately, about 24 hours apart. Patchy cloud
features are visible at most longitudes, and they evolve rapidly from
day to day.}
\label{fig:Hubble_maps}
\end{figure}

Mapping the {\it Hubble} data at 763 nm from 2015 and 2016,
Figure~\ref{fig:Hubble_maps} shows that patchy cloud features are
visible at most longitudes. These features evolve rapidly from day to
day.  In the 2015 data, the dominant feature was the cloud system near
$45^\circ$ S, while in 2016, it is the larger storm complex between 15
and $45^\circ$ S.  Although there is incomplete coverage, there are
still clouds at most longitudes, such that a 763 or 845-nm light curve
is still sensitive to features outside the largest storm system. 
However, if one only considers the brightness in the large bright
cloud systems, compared with the darkest regions, the brightness
varies by about 1.4 magnitudes in 2015 and 1.7 magnitudes in 2016 at
763 nm.  Thus, the amplitudes of the light curves from the {\it
Spitzer} and {\it WISE} filters are consistent with being most
sensitive to the highest clouds against a dark background.  

\section{Summary and Conclusions}

The synergy between Solar System and various brown dwarf and
extrasolar planet cloud decks has been apparent for some time.
\citet{ackerman2001} suggested for example that the L to T type
transition in brown dwarfs might be associated with the appearance of
regions of low cloud opacity, analogous to Jupiter's 5 $\mu$m hot
spots. Many early searches for brown dwarf variability were likewise
motivated by such comparisons. However a major difficulty is that,
while there is a wealth of spatially resolved snapshots of giant
planet atmospheres, there was a dearth of high cadence, full disk
multi-band photometry of any planet in the outer Solar System. This
led \citet{gelino2000}, for example, to synthesize an artificial light
curve of Jupiter from static optical and thermal infrared images. The
last few years has finally seen a new interest in such observations
and a number of ``Solar System planets as point sources" observations
have been obtained by various observatories.

Here we presented full disk observations of an unresolved Neptune by
the {\it Spitzer} Space Telescope and {\it WISE}. Large amplitude
($\sim$ 1 mag) variability is detected, likely arising from high
contrast, optically thick cloud decks that overlie much of the
atmospheric methane opacity. Rotational variability due to bright
clouds as seen in Figure~\ref{fig:Hubble_maps} are consistent with our
$\sim$3 to $\sim$5 $\mu$m photometric signal. A similar result came
from comparing the time series obtained by the {\it K2} mission with
{\it HST} images. Large scale cloud features evolve in height, size,
and overall contrast, and they appear and disappear at different
latitudes. The single {\em Spitzer} visit to the planet reported on
here was insufficient to detect such evolution. A longer, systematic
sampling of the ice giant lightcurves by {\it Spitzer} accompanied by
disk resolved imaging by {\it HST} would provide an outstanding
dataset for comparative studies of weather in substellar objects.

\begin{acknowledgements}
This work is based on observations made with the {\em Spitzer} Space
Telescope, which is operated by the Jet Propulsion Laboratory (JPL),
California Institute of Technology (Caltech), under a contract with
the National Aeronautics and Space Administration (NASA). Support for
this work was provided by NASA through an award issued by
JPL/Caltech. 

This publication makes use of data products from {\it WISE}, which is
a joint project of the University of California, Los Angeles, and
JPL/Caltech, funded by the NASA. This research has made use of the
NASA/IPAC Infrared Science Archive, which is operated by JPL/Caltech,
under contract with NASA.  

This publication makes use of data products from the NASA/ESA Hubble
Space Telescope, under programs GO13937 and GO14492, which is operated
by the Association of Universities for Research in Astronomy, Inc.,
under NASA contract NAS5-26555, with special thanks to the GO14492
science team (M.H.~Wong, A.A.~Simon, I.~de Pater, J.W.~Tollefson,
K.~de Kleer, H.B.~Hammel, S.~Cook, R.~Hueso, A.~S\'anchez-Lavega,
M.~Delcroix, L.~Sromovsky, G.~Orton, and C.~Baranec).

This research was carried out in part at JPL/Caltech under a contract
with NASA and with the support of the NASA Origins of Solar Systems
program via grant 11-OSS11-0074.  

\end{acknowledgements}

\facility{Spitzer}
\facility{K2}
\facility{HST} 

\newpage
\newpage

\end{document}